\documentclass[aps,prl,twocolumn,superscriptaddress,showpacs,floats]{revtex4}

\usepackage{bm}
\usepackage{psfig}

\begin{document}

\title{How to measure squeezing and entanglement of Gaussian states
without homodyning}

\author{Jarom\'{\i}r Fiur\'{a}\v{s}ek}
\affiliation{QUIC, Ecole Polytechnique, CP 165, 
Universit\'{e} Libre de Bruxelles, 1050 Bruxelles, Belgium }
\affiliation{Department of Optics, Palack\'{y} University, 
17. listopadu 50, 77200 Olomouc, Czech Republic}

\author{Nicolas J. Cerf\,}
\affiliation{QUIC, Ecole Polytechnique, CP 165, 
Universit\'{e} Libre de Bruxelles, 1050 Bruxelles, Belgium }

\begin{abstract}
We propose a scheme for measuring the squeezing, purity,
and entanglement of Gaussian states of light that does not require homodyne detection. 
The suggested setup only needs beam splitters and single-photon
detectors. Two-mode entanglement can be detected from coincidences
between photodetectors placed on the two beams.

\end{abstract}


\maketitle

The recent rapid development of quantum information theory has 
largely stimulated research on nonclassical states of light, with the main focus on the generation of entangled states of
light that are required for tasks such as quantum teleportation, 
dense coding, or certain types of quantum  key distribution protocols. 
A particularly promising approach consists in processing 
quantum information with continuous variables \cite{Braunsteinbook}, 
where the quantum information is encoded
into two conjugate quadratures of the quantized mode of the optical field. 
The main advantage of this approach is that many protocols
can be implemented by processing squeezed
light into linear optical interferometers followed by measurements with
highly efficient photodiodes \cite{Braunsteinbook}. 
Such experiments can be described in terms of
Gaussian states which thus play a central role 
in continuous-variable quantum information processing.
In particular, squeezed Gaussian states provide the necessary source 
of entanglement.  The squeezing is usually
observed with the use of a balanced homodyne detector, where the signal beam is
combined with a strong local oscillator (LO) providing a phase reference
\cite{Bachorbook}. The observed quadrature fluctuations depend on the relative phase between the LO and the signal. The maximal squeezing
 of the signal
then corresponds to the minimal observed quadrature variance.
\par

Given that quadrature squeezing is inherently a phase-sensitive phenomenon, one would expect that it may not be possible to determine the squeezing properties
without an external phase reference (LO). In this paper, we show
that, surprisingly, a phase-insensitive device is sufficient
provided that we can \emph{a priori} 
assume that the optical mode is in a Gaussian state. 
The setup we suggest 
consists in beam splitters with variable splitting ratios, phase shifters,
and photodetectors with single-photon sensitivity (e.g.,  avalanche
photodiodes). It can also be extended to estimate the squeezing of multimode
Gaussian states. In particular,
a variation of our setup is capable of measuring the degree of
entanglement of a two-mode Gaussian state, namely the logarithmic
negativity \cite{Eisertthesis,Vidal02}. Besides the determination of
squeezing and entanglement, our setup 
can also be used to measure the purity of Gaussian states \cite{Kim02}.
In addition, the detectors need not be perfect and an efficiency $\eta<1$
can easily be compensated by proper data processing. 
\par

Our scheme works for an arbitrary number of modes $N$ and 
is economic with respect to $N$ in the sense that the number of measured
parameters is  only linear in $N$ while the full tomography of Gaussian states
revealing the whole covariance matrix 
would require the measurement of $\propto N^2$ parameters. In this context,
it is related to several recent proposals on how to directly measure the 
purity, overlap, and entanglement of quantum states without full state
reconstruction \cite{Filip02,Horodecki02,Ekert02,Hendrych03}. It is also
reminiscent of the photon-number distribution measurement scheme
using a photodetector without single-photon resolution
as proposed in Ref. \cite{Mogilevtsev98}.
\par

\paragraph{Preliminaries.} Let us begin with
introducing the necessary notation and definitions. Let
$r=(x_1,p_1,\ldots, x_k,p_k,\ldots, x_N,p_N)$ be the vector of conjugate 
quadratures of $N$ modes which satisfy the canonical commutation relations
$[x_j,p_k]=i\delta_{jk}$. The Gaussian state is fully described
by the vector of mean values $\xi_j=\langle r_j \rangle$ 
and the covariance matrix 
\begin{equation}
\gamma_{jk}=\langle \Delta r_j \Delta r_{k}\rangle +
\langle \Delta r_k \Delta r_j\rangle,
\label{gammadefinition}
\end{equation}
where $\Delta r_j=r_j-\xi_j$.
The quantum state of the optical field can be fully characterized by a
$s$-parametrized quasidistribution which provides phase-space representation of
the quantum state. For our purposes, it is convenient to utilize the Husimi
Q-function. The Q-function of an $N$-mode Gaussian state is the Gaussian distribution \cite{Perinabook}
\begin{equation}
Q(r)=\frac{\pi^{-N}}{ \sqrt{\det (\gamma+I)}} 
\exp \left[ -(r-\xi)^{T}(\gamma+I)^{-1}(r-\xi) \right] ,
\label{Qfunction}
\end{equation}
where $I$ is the identity matrix. 
The squeezing properties do not depend on $\xi$ 
and are fully described by $\gamma$. The maximal observable 
squeezing, i.e. the minimal quadrature variance, is called the generalized squeezing
variance $\lambda$  and can be determined as the minimal eigenvalue of the 
covariance matrix \cite{Simon94},
\begin{equation}
\lambda=\min[\mathrm{eig} (\gamma)].
\label{lambdadefinition}
\end{equation}
The purity of the mixed state with density matrix $\rho$ is defined as 
$\mathcal{P}=\mathrm{Tr}[\rho^2]$. For a Gaussian state with covariance matrix 
$\gamma$, one obtains \cite{Kim02}
\begin{equation}
\mathcal{P}= [\det(\gamma)]^{-1/2}.
\label{purity}
\end{equation}
\par

\begin{figure}[t]
\centerline{\psfig{figure=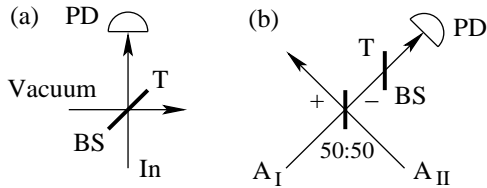,width=0.75\linewidth}}
\caption{Direct measurement of a Gaussian state with the use of
a photodetector PD with single-photon sensitivity, preceeded 
by a beam splitter BS of transmittance $T$. (a) Single-copy scheme
for $\xi=0$. (b) Two-copy scheme for $\xi\ne 0$. }
\end{figure}

\par

\paragraph{Single-mode case.} For the sake of simplicity, 
we first illustrate the procedure on single-mode Gaussian states ($N=1$).
Consider the setup depicted in Fig. 1(a). The input mode impinges on a beam
splitter BS with tunable transmittance $T$, and the output mode 
is measured by a photodetector PD with efficiency $\eta$ that is sensitive to single photons (no single-photon resolution is needed). 
We assume that this realistic detector can be modeled as 
a beam splitter with transmittance $\eta$ followed by an ideal detector that
performs a dichotomic measurement described by the POVM elements 
$\Pi_0=|0\rangle\langle 0|$ and $\Pi_1=\openone - \Pi_0$. 
(In what follows, we assume that the detector is ideal and $\eta$ can be taken
into account by substituting $T\rightarrow \eta T$.) 
The probability of no-click of an ideal detector PD is given by
$P=\mathrm{Tr}[\rho \Pi_0]=\langle 0|\rho|0\rangle=2\pi Q(0)$,
so that inserting $r=0$ into Eq. (\ref{Qfunction}) yields
\begin{equation}
P=\frac{2}{\sqrt{\det (\gamma^\prime+I)}}
\exp\left[- \xi^{\prime T}(\gamma^\prime+I)^{-1}\xi^\prime \right].
\label{probability1}
\end{equation}
where $\gamma^\prime$ and $\xi^\prime$ are, respectively, 
the covariance matrix and the displacement vector of the beam 
impinging on the photodetector.
\par

Suppose that we set the beam splitter transmittance to the value $T_j$.
The covariance matrix $\gamma'$ of the state after passing the beam splitter
reads $\gamma'=T_j\gamma+(1-T_j)I$. Similarly, the coherent signal is damped to
$\xi'=\sqrt{T_j}\xi$. On inserting $\gamma'$ and $\xi'$ 
into Eq. (\ref{probability1}), we obtain 
\begin{equation}
P_{j}=\frac{2}{[\det (\tilde{\gamma}_{T_j})]^{1/2}}
\exp\left(- T_j\xi^{T}\tilde{\gamma}_{T_j}^{-1}\xi \right).
\label{probability}
\end{equation}
where $\tilde{\gamma}_{T_j}=T_j\gamma+(2-T_j)I$.
We thus find that
$P_j$ depends on $T_j$ and four parameters of the state: 
$\det\gamma$, $\mathrm{Tr}(\gamma)$, $g_1=\xi_1^2\gamma_{22}
+\xi_2^2\gamma_{11}-2\gamma_{12}\xi_1\xi_2$, and $g_2=\xi_1^2+\xi_2^2$.
This immediately suggests 
that if we measure $P_j$ for at least four different $T_j$'s,
we might be able to reconstruct the values of these four parameters 
by solving a system of nonlinear equations. 
However, numerical simulations reveal that
the inversion of these highly nonlinear equations
typically leads to extremely large fluctuations 
of the estimated parameters even
for a very large number of measurements for each setting $T_j$. 
\par

Fortunately, in the important case where the displacement vector is zero
($\xi=0$), the scheme provides reliable and well-behaved estimates
of these parameters since formula~(\ref{probability}) 
then simplifies to  
\begin{equation}
\det\tilde{\gamma}_{T_j} =T^2\det(\gamma) 
+T(2-T)\mathrm{Tr}(\gamma)  + (2-T)^2
= 4 P_{j}^{-2}.
\label{linearinversion}
\end{equation}
This results in a system of \emph{linear} 
equations for $\det(\gamma)$ and $\mathrm{Tr}(\gamma)$. 
If measurements for two different transmittances $T_1$ and
$T_2$ are performed and the observed probabilities of no-click
are $P_1$ and $P_2$, then the
system of Eqs. (\ref{linearinversion}) can easily be solved and yields
\begin{equation}
\mathrm{Tr}(\gamma)=\frac{2}{T_2-T_1}\left(\frac{T_2}{T_1P_1^2}
-\frac{T_1}{T_2P_2^2}\right)+2-\frac{2}{T_1}-\frac{2}{T_2},
\label{trgammasolution}
\end{equation}
\begin{equation}
\det(\gamma)=\frac{2}{T_1-T_2}\left(\frac{2-T_2}{T_1P_1^2}-\frac{2-T_1}{T_2P_2^2}
\right)+\frac{(2-T_1)(2-T_2)}{T_1T_2}.
\label{detgammasolution}
\end{equation}
Let us investigate what can be extracted from
the knowledge of $\mathrm{Tr}(\gamma)$ and $\det(\gamma)$. 
As noted above, the squeezing properties of the
Gaussian state, namely the generalized squeezing variance $\lambda$, can be
determined from the eigenvalues of $\gamma$, {\it cf.} 
Eq.~(\ref{lambdadefinition}).
For a single-mode state, $\gamma$ is symmetric $2\times 2$ matrix and its 
eigenvalues can be expressed in terms of $\det(\gamma)$ and $\mathrm{Tr}(\gamma)$, 
which are both determined by the present method. We find that
\begin{equation}
\lambda=\frac{1}{2}
\left[\mathrm{Tr}(\gamma)-\sqrt{\mathrm{Tr}^2(\gamma)-4\det(\gamma)}\right].
\label{lambdasinglemode}
\end{equation}
Moreover, since our method provides an estimate of $\det{\gamma}$, we can also 
determine the purity from Eq. (\ref{purity}). 
\par

If $\xi\ne 0$, our scheme is still usable provided that we can perform
a collective measurement on two copies of the state, as depicted in Fig. 1(b). 
The two input modes $A_{I}$ and $A_{II}$ prepared in identical
Gaussian state interfere on a balanced beam splitter. In the Heisenberg picture,
the annihilation operators of the output modes $A_{+}$ and $A_{-}$ are linear
combinations of those of the input modes, $a_{\pm}=2^{-1/2}(a_{I}\pm a_{II})$.
The covariance matrix of mode $A_{-}$  is equal to
$\gamma$ \cite{Browne03} but the coherent signal in $A_{-}$ vanishes
due to the destructive interference, $\xi_{-}=0$. As shown in Fig. 1(b), the mode $A_{-}$ is subsequently sent to a direct measurement setup
identical to that shown in Fig. 1(a). 
\par

\paragraph{Multimode case.} We now extend this procedure 
to multimode Gaussian states. A reliable operation again requires
two copies $A_{I}$ and $A_{II}$. Essentially, we use 
in parallel $N$ setups such as shown in Fig. 1(b). 
We combine each pair of modes $A_{I,k}$
and $A_{II,k}$, with $k=1,\ldots,N$, on a balanced beam splitter. 
Each ``minus'' mode $A_{-,k}$, with $\xi=0$, is then sent 
to a beam splitter of transmittance $T_j$ followed by a 
photodetector. We measure the probability 
$P_j=2^N/(\det\tilde{\gamma}_{T_j})^{1/2}$ that none of the 
$N$ detectors clicks. For an $N$-mode state, the determinant of 
$\tilde{\gamma}_{T}=T\gamma+(2-T)I$ can be expanded as
\begin{equation}
\det(\tilde{\gamma}_T)=\sum_{n=1}^{2N} T^n(2-T)^{2N-n}f_n(\gamma)+(2-T)^{2N},
\label{detgammaexpansion}
\end{equation}
where $f_{n}(\gamma)$ is an homogeneous polynomial of $n$th order in the matrix
elements of $\gamma$, {\it e.g.}, $f_{2N}(\gamma)=\det(\gamma)$ and 
$f_1(\gamma)=\mathrm{Tr}(\gamma)$. 
The probability  $P_j$ thus
depends on $T_j$ and the $2N$ parameters $f_n(\gamma)$. 
If we measure $P_j$ for $2N$
(or more) different transmittances $T_j$'s, then we can determine the 
parameters $f_n$ of the Gaussian state by solving a system of 
linear equations
\begin{equation}
\sum_{n=1}^{2N} T_j^n(2-T_j)^{2N-n}f_n(\gamma)=2^{2N} P_{j}^{-2}
-(2-T_j)^{2N}.
\end{equation}
Once we know $f_n(\gamma)$, we can determine the generalized
squeezing variance $\lambda $ as the smallest root of the characteristic polynomial
$\det(\lambda I -\gamma)=0$. It can be seen from Eq. (\ref{detgammaexpansion}) 
that the parameters $f_j(\gamma)$ are the coefficients of this characteristic 
polynomial, and we have
\begin{equation}
\lambda^{2N} + \sum_{k=0}^{2N-1} \lambda^{k} (-1)^k f_{2N-k}(\gamma)=0.
\end{equation}
We can also determine the purity of the $N$-mode Gaussian state 
from $f_{2N}(\gamma)$ with the help of formula (\ref{purity}).
\par

\paragraph{Entanglement detection.}
In the context of quantum information processing with continuous variables,
the entanglement properties of Gaussian states deserve particular attention. 
It has been shown that a two-mode Gaussian state is separable 
iff it has a positive
partial transpose \cite{Duan00,Simon00}. This property can easily be checked 
if one knows the covariance matrix 
\begin{equation}
\gamma_{AB}=\left(
\begin{array}{cc}
\gamma_{A} & \sigma_{AB} \\
\sigma_{AB}^T & \gamma_{B} 
\end{array}
\right)
\end{equation}
of the bipartite state, where $\gamma_{A}$ and $\gamma_{B}$ are the covariance 
matrices of modes $A$ and $B$, respectively, while $\sigma_{AB}$ captures 
the intermodal correlations.  Moreover, analytical formulas for
several entanglement monotones that measure the entanglement
of Gaussian states have been given in the literature
\cite{Vidal02,Giedke03}. 
A particularly simple formula has been obtained for the logarithmic negativity
$E_{\mathcal{N}}$ of an arbitrary Gaussian state. 
To calculate $E_{\mathcal{N}}$, we must determine
the symplectic spectrum of the covariance matrix of the partially transposed 
state $\rho_{AB}^{T_A}$. As shown in Ref. \cite{Vidal02}, the symplectic 
eigenvalues are
the two positive roots $\zeta_1 \geq \zeta_2 >0$ of the biquadratic equation
\begin{equation}
\zeta^4-(\det\gamma_{A}+\det\gamma_{B}-2\det\sigma_{AB})\zeta^2+\det\gamma_{AB}=0.
\label{symplecticequation}
\end{equation}
The solution of Eq.~(\ref{symplecticequation}) yields 
\begin{equation}
\zeta_2^2=\frac{1}{2}\left(D-\sqrt{D^2-4\det\gamma_{AB}}\right),
\label{zetatwo}
\end{equation}
where $D=\det\gamma_{A}+\det\gamma_{B}-2\det\sigma_{AB}$.
The two-mode Gaussian state is entangled if and only if $\zeta_2<1$. 
In this case, we have
$E_{\mathcal{N}}=-\log(\zeta_2)$, while $E_{\mathcal{N}}=0$ otherwise.
The condition $\zeta_2<1$  implies the necessary and sufficient 
entanglement condition
$D>1+\det\gamma_{AB}$ \cite{Simon00,Giedke01}, which explicitly reads,
\begin{equation}
\det\gamma_{A}+\det\gamma_{B}-2\det\sigma_{AB}>1+\det\gamma_{AB}.
\label{entanglementcriterion}
\end{equation}
\par

With the use of the method proposed in the present paper we can measure 
$\det\gamma_{AB}$, $\det\gamma_{A}$, and $\det\gamma_{B}$. 
An upper bound on $\det\sigma_{AB}$ in terms of these determinants 
can be derived from the condition that 
the symplectic eigenvalues $\tilde{\zeta}_j$ of the covariance matrix $\gamma_{AB}$ 
must be  greater or equal to one \cite{Giedke01}. The lower 
eigenvalue $\tilde{\zeta}_2$ is given by Eq.~(\ref{zetatwo}), where $D$ is replaced with 
$D'=\det\gamma_{A}+\det\gamma_{B}+2\det\sigma_{AB}$. The condition
$\tilde{\zeta}_2^2\geq 1$ yields 
\begin{equation}
2\det\sigma_{AB}\leq \det\gamma_{AB}+1-\det\gamma_A-\det\gamma_B.
\label{sigmabound}
\end{equation}
This, in turn, implies an upper bound on the lower symplectic eigenvalue 
$\zeta_2$ of the covariance matrix of $\rho_{AB}^{T_A}$. On inserting the 
upper bound on $2\det\sigma_{AB}$ given by Eq. (\ref{sigmabound}) 
into Eqs. (\ref{zetatwo}) and (\ref{entanglementcriterion})
we find that $\zeta_2<1$, so that the state is entangled when
\begin{equation}
\det\gamma_A+\det\gamma_B>1+\det\gamma_{AB}
\label{detinequality}
\end{equation}
holds. Inequality (\ref{detinequality}) is thus a sufficient condition 
for entanglement, but it is not
necessary as some Gaussian entangled states are not detected by this test.
The main advantage of this test is that all determinants 
appearing in Eq. (\ref{detinequality})
can be determined by \emph{local} measurements supplemented with classical 
communication between $A$ and $B$. Moreover, if we can {\it a priori} 
assume that  $\xi=0$, 
then measurements on a single copy of $\rho_{AB}$ suffice.
\par

If we now want to exactly determine $E_{\mathcal{N}}$,
we also need a scheme to measure $\det\sigma_{AB}$. 
This can be accomplished provided that we can perform joint measurements on
several copies of the state $\rho_{AB}$. Unlike the previous one,
this scheme requires joint nonlocal measurements on modes $A$ and $B$, and it
involves several steps as schematically illustrated in Fig. 2. 
Using the scheme of Fig. 2(b), we can 
measure the  determinants of the covariance matrices $\gamma_{+}$ and $\gamma_{-}$ of modes $A_{+}$  and $A_{-}$ 
that are linear combinations of the modes
$A$ and $B$, $a_{\pm}=(a\pm b)/\sqrt{2}$. After a simple algebra, we find that
\begin{equation}
\det(\sigma_{AB}+\sigma_{AB}^T)=
2\det\gamma_{+}+2\det\gamma_{-}-\det(\gamma_{A}+\gamma_{B}).
\label{gammaplusminussum}
\end{equation}
In order to determine $\det(\gamma_A+\gamma_B)$, we have to carry out 
a joint measurement on two independent copies of the two-mode state,
as depicted in Fig. 2(c). 
By mixing the modes $A_1$ and $B_2$ on a balanced beam splitter, we
prepare an output single-mode state with covariance matrix
$\gamma_{A+B}\equiv (\gamma_{A}+\gamma_{B})/2$. 
Recall that the modes $A_1$ and $B_2$ belong to two
independent copies of the two-mode state $\rho_{AB}$,
hence $A_1$ and $B_2$ are uncorrelated.  
After the measurement of $\det\gamma_{+}$, $\det\gamma_{-}$,
and $\det\gamma_{A+B}$, we calculate $\det(\sigma_{AB}+\sigma_{AB}^T)$ 
from Eq. (\ref{gammaplusminussum}). It holds that
$\det(\sigma_{AB}+\sigma_{AB}^T)\leq 4\det\sigma_{AB}$ and the equality is achieved
when $\sigma_{AB}$ is symmetric. The matrix $\sigma_{AB}$ can be brought to a symmetric
form by applying a local phase shift $\exp(i\phi b^\dagger b)$ to the mode $B$
using the phase shifter PS in Fig. 2. This transforms $\sigma_{AB}$ to
$\sigma_{AB} U(\phi)$ where 
\[ 
U(\phi)=\left(
\begin{array}{cc}
\cos\phi & \sin\phi \\
-\sin\phi & \cos\phi
\end{array}
\right).
\]
To determine the phase shift $\phi$ that symmetrizes $\sigma_{AB}$,
we measure $y_{\phi}\equiv \det[\sigma_{AB}U(\phi)+U^T(\phi)\sigma_{AB}^T]$
for three different phases $\phi=0$, $\pi/4$, and $\pi/2$. It can be shown that 
\[
y_{\phi}=y_0\cos^2\phi+y_{\pi/2}\sin^2\phi+\tilde{y}_{\pi/4}\sin(2\phi),
\]
where $\tilde{y}_{\pi/4}=y_{\pi/4}-(y_0+y_{\pi/2})/2$. The value of
$\det\sigma_{AB}$ can be found as the maximum of $y_\phi$ over $\phi$ which yields
\[
\det\sigma_{AB}=\frac{1}{8}\left[
y_0+y_{\pi/2}+\sqrt{(y_0-y_{\pi/2})^2+4\tilde{y}_{\pi/4}^2} \,
\right].
\]
This finally provides all the information required for the exact 
calculation of the logarithmic negativity $E_{\mathcal{N}}$.
\par

\begin{figure}[!t!]
\centerline{\psfig{figure=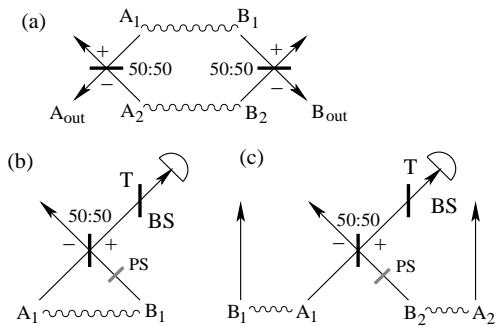,width=0.75\linewidth}}
\caption{Direct measurement of the intermodal correlations 
$\det\sigma_{AB}$ of a two-mode Gaussian state. 
(a) Preparation of states with zero displacement $\xi=0$
needed for steps (b) and (c).
(b) Measurement of $\det\gamma_{+}$ ($\det\gamma_{-}$ is measured similarly).
(c) Measurement of $\det\gamma_{A+B}$.}
\end{figure}

In summary, we have proposed a scheme for the direct measurement
of the squeezing, purity, and entanglement of Gaussian states 
that does not require homodyne detection 
but only needs beam splitters and photodetectors with single-photon
sensitivity. 
The scheme generally requires joint measurements on two copies of the
state, but single-copy measurements suffice if it is 
{\it a priori} known that the mean (or coherent) values 
of the quadratures vanish, which is, {\it e.g.}, the case of the
squeezed and entangled states generated by spontaneous parametric
downconversion. We have shown that, based on Eq. (\ref{detinequality}),
the present method can be used to assess the entanglement 
of $1\times 1$ Gaussian states by means of local measurements, 
without employing 
any local oscillator or interferometric schemes. Given the simplicity of the 
suggested setup, the prospects for an 
experimental realization in a near future look very good.

{\it Note added:} The sufficient condition on entanglement 
[Eq.~(\ref{detinequality})] has recently and independently been
derived by Adesso \emph{et al.} \cite{Adesso03}. Besides the lower bound on $E_\mathcal{N}$ linked to Eq.~(\ref{detinequality}),
an upper bound on $E_{\mathcal{N}}$ expressed in terms 
of the determinants of $\gamma_A$, $\gamma_{B}$,
and $\gamma_{AB}$ was also derived in Ref. \cite{Adesso03}. 
Remarkably, these two bounds are
typically very close to each other, so the knowledge of the determinants 
of the covariance matrices provides quite precise quantitative information
on the entanglement, making the direct measurement procedure particularly
powerful.

We acknowledge financial support from the Communaut\'e Fran\c{c}aise de
Belgique under grant ARC 00/05-251, from the IUAP programme of the Belgian
government under grant V-18, from the EU under projects RESQ
(IST-2001-37559) and CHIC (IST-2001-32150). JF also acknowledges support
from the  grant LN00A015  of the Czech Ministry of Education.

\end{document}